\documentclass[preprint,3p, twocolumn]{elsarticle}
\usepackage[fleqn]{amsmath}
\usepackage{amssymb,amsthm}
\usepackage{mathtools}
\usepackage{mathrsfs}
\usepackage{bm}
\usepackage{slashed}	%feynman slashed symbols
\usepackage{cancel}
\usepackage{soul}
\usepackage{color}
\usepackage{xcolor}
\usepackage[pdftex,hypertexnames=false,linktocpage=true]{hyperref}
\hypersetup{colorlinks=true,linkcolor=red,anchorcolor=darkgreen,citecolor=green!50!blue,filecolor=darkgreen,urlcolor=green,
bookmarksnumbered=true,
pdfview=FitB
}
\usepackage{lineno}
\biboptions{sort&compress}
\usepackage{verbatim}
\usepackage{fullpage}
\usepackage{graphicx}
\usepackage{subfig} 	% a set of figures
\usepackage{relsize}	%rescalable math
\usepackage{float}
\usepackage{array}
\usepackage{booktabs}
\usepackage{arydshln}
\usepackage{multirow}
\makeatletter

\newcommand{\Rmnum}[1]{\expandafter\@slowromancap\romannumeral #1@}
\makeatother
\colorlet{darkgreen}{green!60!black}
\colorlet{brightyellow}{yellow!75!red}
\colorlet{orange}{red!50!yellow}
\colorlet{darkblue}{blue!60!black}
\colorlet{darkred}{red!80!black}
\colorlet{greenblue}{green!50!blue}

\begin{document}
\begin{frontmatter}

%title and authors
\title{Pion spectroscopy and dynamics using the holographic light-front Schr\"odinger equation and the 't Hooft equation}

\author[mau]{Mohammad Ahmady}
\ead{mahmady@mta.ca}
\author[imp,cas]{Satvir Kaur}
\ead{satvir@impcas.ac.cn}
\author[imp,cas]{Chandan Mondal}
\ead{mondal@impcas.ac.cn}
\author[au]{Ruben Sandapen}  
\ead{ruben.sandapen@acadiau.ca}
\address[mau]{Department of Physics, Mount Allison University,  Sackville, New Brunswick,  E4L 1E6, Canada.}
%\address[nitj]{}
\address[imp]{Institute for Modern Physics, Chinese Academy of Sciences, Lanzhou-730000, China} 
\address[cas]{School of Nuclear Science and Technology, University of Chinese Academy of Sciences, Beijing 100049, China}
%\address[impcas]{CAS Key Laboratory of High Precision Nuclear Spectroscopy, Institute of Modern Physics, Chinese Academy of Sciences, Lanzhou 730000, China}
\address[au]{Department of Physics, Acadia University, Wolfville, Nova Scotia, B4P 2R6, Canada.}
%\address[ptu]{ Department of Physical Sciences, 
%	I K Gujral Punjab Technical University,
%	Kapurthala-144603, Punjab, India.}
%%abstract

\begin{abstract}
We show that the holographic Schr\"odinger equation of light-front chiral QCD, together with 
	the 't Hooft equation of (1+1)-dimensional QCD in the large $N_c$ limit, can simultaneously describe pion spectroscopy as well as the pion decay constant, charge radius, electromagnetic form factor,  photon-to-pion transition form factor, Parton Distribution Function (PDF) and Distribution Amplitude (DA). Furthermore, the chiral-limit constraints, as encoded in the Gell-Mann-Oakes-Renner (GMOR) relation, are satisfied. 
	\end{abstract}
\begin{keyword}
 Light-front holographic QCD \sep 't Hooft equation \sep Pions \sep Longitudinal dynamics \sep Chiral symmetry breaking \sep Confinement
\end{keyword}

\end{frontmatter}

\section{Introduction} 
Hadrons are bound states of quarks held together by the strong interaction described by Quantum Chromodynamics (QCD).  Due to the nonperturbative behavior of the strong interaction at large distances, direct calculations of the hadronic properties based on the underlying QCD interactions is only possible through numerical lattice QCD simulations.  Alternatively, one can compute hadronic observables using QCD models that capture the main features of the strong interaction such as chiral symmetry breaking and confinement. The pion is an ideal particle to investigate these two features since it is both a QCD bound state and a Nambu-Goldstone boson of chiral QCD. More generally, for a pseudoscalar meson of mass $M_P$, decay constant $f_P$, made up of quarks with mass $m_q$, chiral symmetry breaking and confinement are encapsulated in the Gell-Mann-Oakes-Renner (GMOR) relation \cite{GMOR}
\begin{equation}
	M_P^2 f_P^2=-2\langle q\bar{q} \rangle m_q + \mathcal{O} (m_q^2)\,,
	\label{GMOR}
\end{equation}
where $\langle q\bar{q} \rangle$ is interpreted, in light-front QCD, as the in-meson quark condensate \cite{Brodsky:2012ku}. While  Eq.~\eqref{GMOR} holds for the ground state pion and its excited states, only the ground state pion is a Nambu-Goldstone boson,  with nonvanishing decay constant in the chiral limit \cite{Li:2016dzv}. Indeed,  Eq.~\eqref{GMOR} implies that $M^2_\pi \propto m_q$, provided that $f_\pi$ does not vanish in the chiral limit. On the other hand,  for the excited pions, $\pi^\prime\equiv \pi(1300)$ and $\pi^{\prime \prime}\equiv \pi(1800)$, Eq.~\eqref{GMOR} implies that  $f^2_{\pi^{\prime,\prime \prime}} \propto m_q$, provided that their masses do not vanish in the chiral limit. For small but nonzero quark masses, we therefore expect both the masses of the ground state pions and the decay constants of excited pions to be suppressed: $M_\pi \ll M_{\pi^{\prime,\prime \prime}}$ and $f_{\pi^{\prime, \prime \prime}} \ll f_\pi$.

Holographic light-front QCD (hLFQCD) is formulated in the chiral limit of light-front QCD, where there exists an exact correspondence  between strongly coupled (1+3)-dimensional light-front QCD and weakly interacting string modes in (1+4)-dimensional anti-de-Sitter (AdS) space. For a review of hLFQCD, see Ref. \cite{Brodsky:2014yha}. For a meson, the light-front wavefunction, $\Psi(x,\mathbf{b}_\perp )$, encodes a Lorentz-invariant description of the strong interaction between the quark and antiquark in terms of the momentum fraction $x=k^+/P^+$, where $k^+$ and $P^+$ are the light-front momenta of the quark and the meson respectively, and $\mathbf{b}_\perp =b_\perp e^{i\varphi}$ is the transverse quark-antiquark separation. To connect with AdS space, a holographic variable,    
\begin{equation}
\boldsymbol{\zeta}=\sqrt{x(1-x)} \mathbf{b}_\perp =\zeta e^{i\varphi} \;,
\label{zeta}
\end{equation}
is introduced, and the wavefunction is written in a factorized form in $x$, $\zeta$ and $\phi$ variables:   
\begin{equation}
	\Psi (x, \zeta, \varphi)= \frac{\phi (\zeta)}{\sqrt{2\pi \zeta}} e^{i L \varphi} X(x) \;,
	\label{full-mesonwf}
\end{equation}
where $\phi (\zeta)$ and $X(x)=\sqrt{x(1-x)} \chi(x)$ are referred to as the transverse and longitudinal modes respectively.  We normalize Eq.~\eqref{full-mesonwf} using
\begin{equation}
	\int \mathrm{d}x \mathrm{d}^2 \mathbf{b}_\perp |\Psi(x, \mathbf{b}_\perp)|^2 =1,
\end{equation}
i.e. we assume only the leading valence Fock sector in the meson. 

Even though separating the transverse and longitudinal dynamics is natural on the light-front~\cite{Brodsky:1997de}, one can quantify this separability by computing the entanglement entropy between the transverse and the longitudinal d.o.f~\cite{Eisert:2008ur}.  In Ref~\cite{Li:2017mlw}, a model calculation of this entanglement entropy for charmonium states shows that it is generally small for the ground states. Here, we shall explore the consequences of such a separation for the pion and its excited states. Agreement with the experimental data provides \textit{\`a posteriori} support for this assumed separability.  

\begin{figure}[hbt!]
	\begin{center}
		\includegraphics[width=0.9\linewidth]{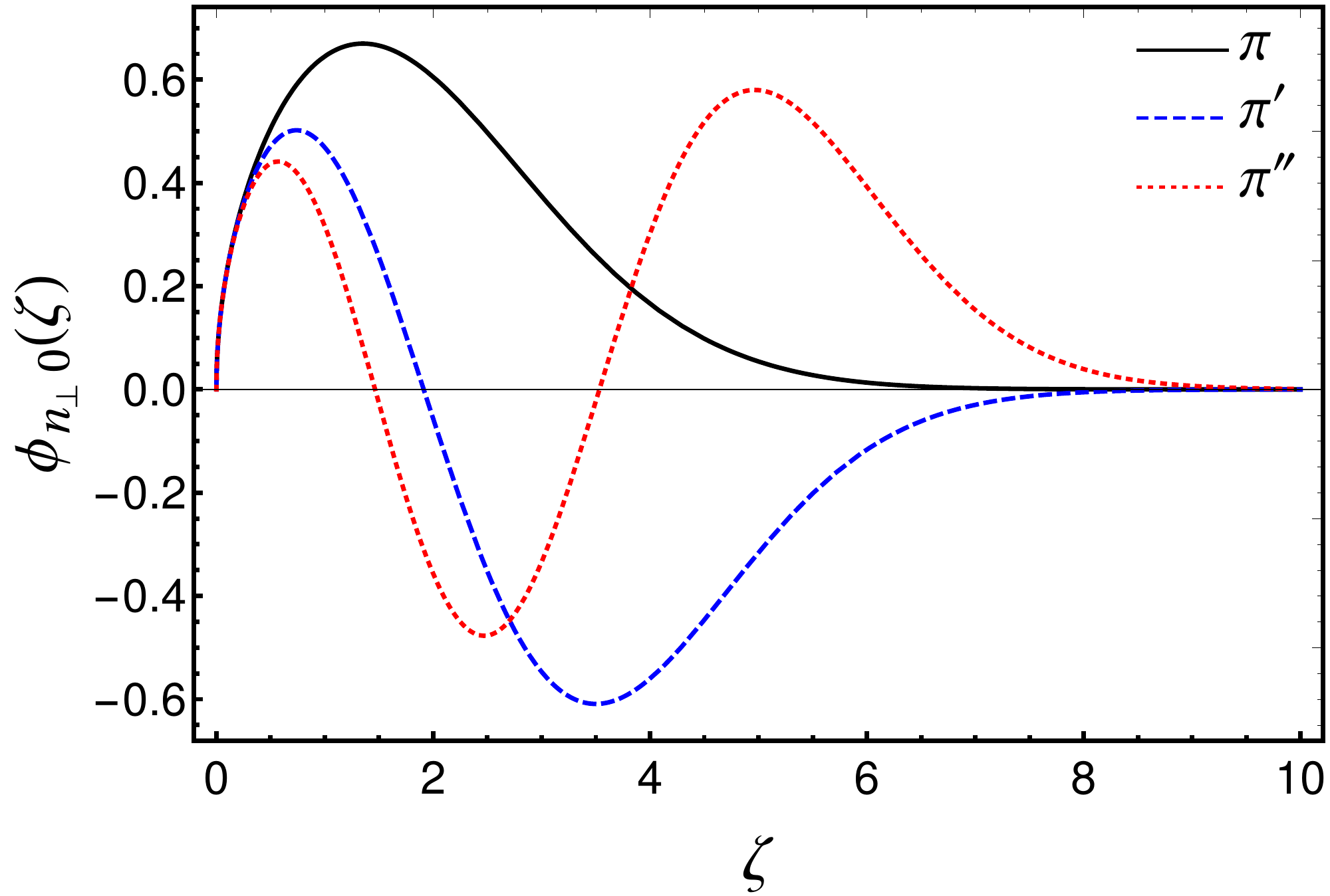}
		\caption{The transverse modes, $\phi (\zeta )$, as function of the holographic variable, $\zeta$, for $\pi$, $\pi^\prime$ and $\pi^{\prime \prime}$ in GeV-based units.}
		\label{tm}
	\end{center}
\end{figure}

In hLFQCD, only the transverse mode is dynamical, being generated by the holographic  Schr\"{o}dinger-like equation~\cite{Brodsky:2006uqa,deTeramond:2005su,deTeramond:2008ht,Brodsky:2014yha}, 
\begin{equation}
	\left(-\frac {\mathrm{d}^2}{\mathrm{d} \zeta^2}+\frac{4L^2-1}{4 \zeta^2}+U^{\mathrm{LFH}}_\perp(\zeta)\right)\phi(\zeta)= M_\perp^2 \phi(\zeta) \;,
	\label{SEq}
\end{equation}
where the transverse confinement potential is given by
\begin{equation}
	U_\perp^{\mathrm{LFH}}(\zeta)=\kappa^4 \zeta^2 + 2\kappa^2(J-1) \;,
	\label{U-LFH}
\end{equation}
with $J=L+S$ ($L\equiv |L^{\mathrm{max}}_z|$) being the total angular momentum of the meson. The analytical form of Eq.~\eqref{U-LFH} is uniquely fixed~\cite{Brodsky:2013ar} by the underlying conformal symmetry and a holographic mapping to $\mathrm{AdS}_5$ where $\zeta$ maps onto the fifth dimension of AdS space. The emerging mass scale, $\kappa$, simultaneously sets the confinement scale and generates meson masses in the chiral limit. With Eq.~\eqref{U-LFH}, Eq.~(\ref{SEq}) can be solved analytically, yielding
\begin{equation}
	M_{\perp}^2(n_\perp , J, L)=4\kappa^2\left(n_\perp + \frac{J+L}{2}\right) \;,
	\label{MTM}
\end{equation}
and 
\begin{equation}
	\phi_{n_\perp L}(\zeta )\propto \zeta^{1/2+L}\exp\left(\frac{-\kappa^2\zeta^2}{2}\right)L_{n_\perp}^L(\kappa^2\zeta^2)\; .
	\label{TMWF}
\end{equation}
A crucial prediction of Eq.~\eqref{MTM} is that the lowest-lying bound state, with $n_\perp =L=S=0$, is massless. This is naturally identified as the pion which is expected to be massless in the chiral limit of QCD. On the other hand, Eq.~\eqref{MTM} predicts the masses of radially excited pions, with $n_\perp=1, L=0$ and $n_\perp=2, L=0$, to be $M_{\pi^\prime}=4\kappa^2$ and $M_{\pi^{\prime \prime}}=8\kappa^2$ respectively, consistent with excited pions not being Nambu-Goldstone bosons in chiral QCD. The resulting transverse wavefunctions for the pion family, $\pi$, $\pi^\prime$ and $\pi^{\prime \prime}$, are shown in Fig.~\ref{tm}, with the characteristics nodes for the excited states.

On the other hand, the longitudinal mode, $\chi(x)$, is not dynamical in hLFQCD, i.e. $\chi(x)=1$ and $X(x)=\sqrt{x(1-x)}$. This is explicitly verified by the holographic mapping of the electromagnetic (or gravitational) pion form factor in physical spacetime and $\mathrm{AdS}_5$~\cite{Brodsky:2007hb,Brodsky:2008pf}. Inserting Eq.~\eqref{TMWF}  in Eq.~\eqref{full-mesonwf} results in the chiral-limit holographic light-front wavefunction for the pion:
\begin{equation}
	\Psi^\pi_{m_q=0} (x,\zeta^2) \propto \sqrt{x(1-x)} \exp\left(\frac{-\kappa^2\zeta^2}{2}\right)\;.
	\label{pichiralwf} 
\end{equation}
A two-dimensional Fourier transform yields 
\begin{align}
	\Phi^\pi_{m_q=0}(x, k^2_\perp )  &\propto \frac{1}{\sqrt{x(1-x)}}\nonumber\\&\times\exp\left(-\frac{k_\perp^2}{2\kappa^2 x(1-x)}\right)\,,
		\label{FT}
\end{align}
where  $\mathbf{k}_\perp$, the transverse momentum of the quark, is the Fourier conjugate of $\mathbf{b}_\perp$. Similarly, the chiral-limit holographic wavefunctions for $\pi^\prime$ and $\pi^{\prime \prime}$ are given by
\begin{align}
\Phi_{m_q=0}^{\pi^\prime}(x, k_\perp^2) &\propto \frac{\left({k}_\perp^2 - x(1-x) \kappa^2 \right)}{{x^{3/2}(1-x)^{3/2}}} \nonumber\\
&\times \exp\left(-\frac{{k}^2_\perp}{2 \kappa^2 x(1-x)} \right)\;,  \label{2S-trans}
\end{align}
and
\begin{align}
&	\Phi_{m_q=0}^{\pi^{\prime \prime}}(x,{k}_\perp^2)\nonumber\\ & \propto  \frac{\left({k}_\perp^4 - 4 {k}_\perp^2 x(1-x) \kappa^2 +2 x^2 (1-x)^2 \kappa^4 \right) }{x^{5/2}(1-x)^{5/2}} \nonumber \\
	&\times \exp\left(-\frac{{k}^2_\perp}{2 \kappa^2 x(1-x)} \right)\,,  \label{3S-trans} 
\end{align}
respectively.	

To move away from the chiral limit, Brodsky and de T\'eramond (BdT) suggested a prescription based on the observation that the chiral-limit of invariant mass of quark-anti-quark pair,
\begin{equation}
	{\cal M}^2_{q\bar q}=\frac{k_\perp^2 +(1-x)m_q^2+xm_{\bar q}^2}{x(1-x)}\; ,
\end{equation}
appears in Eq.~\eqref{FT}, i.e.
\begin{align}
	\Phi^\pi_{m_q=0}(x,k^2_\perp ) &\propto \frac{1}{\sqrt{x(1-x)}}\nonumber\\&\times\exp\left(-\frac{{\cal M}^2_{q\bar q}\mid_{m_q=0}}{2\kappa^2}\right)\;.
	\label{WFIM}
\end{align}
Restoring the quark mass dependence of $\mathcal{M}_{q\bar{q}}$, the Brodsky-de T\'eramond prescription implies that the longitudinal mode becomes~\cite{Brodsky:2008pg}:
\begin{align}
	X(x) &= \sqrt{x(1-x)}\nonumber\\&\times\exp\left(-\frac{(1-x)m_q^2+xm_{\bar q}^2}{x(1-x)}\right)\;.
	\label{bda}
\end{align}
The bound-state mass eigenvalue consequently receives a first-order correction given by
\begin{equation}
	\Delta M^2 = \int \frac{\mathrm{d} x}{x(1-x)} X^2(x) \left(\frac{m_q^2}{x}+\frac{m_{\bar{q}}^2}{1-x} \right)\;.
	\label{massshift}	
	\end{equation}
There are two shortcomings with the above prescription. First, it implies that~\cite{Li:2021jqb}
\begin{equation}
	M_{\pi}^2=\Delta M^2 \propto 2m_q^2 (\ln(\kappa^2/m_q^2)-\gamma_E)\; ,
	\end{equation}
where $\gamma_E=0.577216$ is the Euler's constant, in contradiction with GMOR prediction, $M_{\pi}^2\propto m_q$.  Second, the longitudinal mode, given by  Eq. \eqref{bda}, with no nodes, is also assumed for $\pi^\prime$ and $\pi^{\prime \prime}$. Consequently, this does not lead to a suppression of the decay constants of $\pi^\prime$ and $\pi^{\prime \prime}$.  

References~\cite{Ahmady:2016ufq,Ahmady:2018muv} use Eq.~\eqref{WFIM}, i.e. a non-dynamical longitudinal mode, together with a spin wavefunction, in order to describe pion observables other than the spectroscopic data. On the other hand, Ref.~\cite{Ahmady:2021lsh} considers longitudinal dynamics, generated by the 't Hooft Equation, in order to describe the full meson spectrum but does not predict pion dynamics. Our goal here is to simultaneously describe the pion spectroscopy and dynamics. To do so, we use the longitudinal mode generated by the 't Hooft Equation, as in Ref.~\cite{Ahmady:2021lsh}, but we shall also provide parameter-free predictions for the pion observables considered in Ref.~\cite{Ahmady:2018muv}. At the same time, we show that all the chiral-limit constraints encoded by the GMOR relation, are satisfied. Note that there has been much recent interest in the inclusion of longitudinal dynamics in hLFQCD~\cite{Li:2021jqb,deTeramond:2021yyi,Lyubovitskij:2022rod,Weller:2021wog,Ahmady:2021lsh,Rinaldi:2022dyh}.

 \section{Longitudinal dynamics via the '\lowercase{t} Hooft equation}
 
Starting with the QCD Lagrangian in $(1+1)$-dim in the $N_c \gg 1$ approximation, 't Hooft derived a Schr\"odinger-like equation for the meson given by~\cite{tHooft:1974pnl}
 
 \begin{align}
 	&\left(\frac{m_q^2}{x}+\frac{m_{\bar{q}}^2}{1-x}\right)\chi(x)\nonumber\\& \quad\quad\quad\quad \quad \quad+ U_\parallel (x) \chi(x)=M^2_\parallel \chi(x) \;,
 	\label{tHooft}
 \end{align}
where 
\begin{equation}
	U_\parallel(x)\chi(x)=\frac{g^2}{\pi} \mathcal{P} \int {\rm d}y \frac{\chi(x)-\chi(y)}{(x-y)^2}\;,
	\label{tHooft-potential}
\end{equation}
with $g$ being the longitudinal  confinement scale and $\mathcal{P}$ denoting the Cauchy principal value. Unlike the holographic light-front Schr\"odinger Equation, the 't Hooft does not admit analytical solutions and has to be solved numerically. Here we do so using the matrix method given in Ref.~\cite{Chabysheva:2012fe}. 

Using both the holographic Schr\"odinger Equation together with the 't Hooft Equation, the meson mass is then given by 
\begin{align}
	M^2(n_\perp ,n_\parallel ,J, L)&= 4\kappa^2\left(n_\perp + \frac{J+L}{2}\right)\nonumber\\
	&+ M_\parallel^2(n_\parallel , m_q, m_{\bar q}, g)\; .
	\label{totalmass}
\end{align}

\begin{figure}[hbt!]
	\begin{center}
		\includegraphics[width=0.9\linewidth]{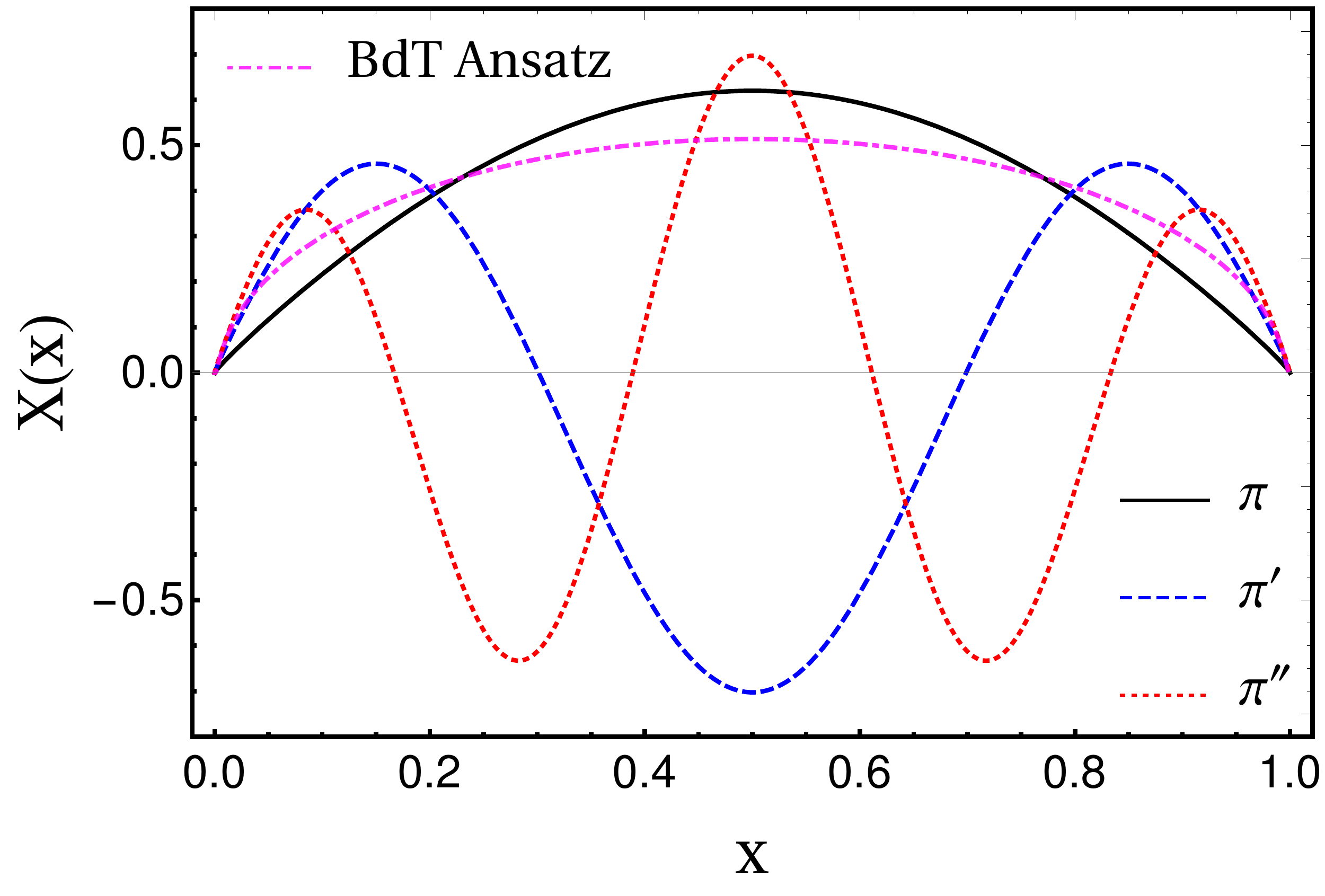}
		\caption{Longitudinal mode $X(x)$ for $\pi$ (solid-black), $\pi^\prime$ (dashed-blue) and $\pi^{\prime \prime}$ (dotted red). For comparison, the Brodsky-de T\'eramond ansatz, Eq.~\eqref{bda}, is shown as the dot-dashed magenta curve.}
		\label{lm}
	\end{center}
\end{figure}

Figure \ref{lm} shows our numerical results for the dynamical longitudinal mode of the ground-state pion compared to those of the excited states $\pi^\prime$ and $\pi^{\prime\prime}$.  We highlight that the dynamical longitudinal modes of the excited pions feature an increasing number of nodes (equal to $n_\parallel$) in contrast to the nondynamical mode,  Eq.~\eqref{bda}, resulting from Brodsky and de T\'eramond prescription. 
%%%%%%%%%%%%%%%%%%%%%%%%%%
\section{Predictions}
%%%%%%%%%%%%%%%%%%%%%%%%%%
We now fix the only $3$ free parameters: the longitudinal and transverse 
confinement scales, $g$ and $\kappa$, and the light quark mass, $m_{u/d}$, in order to fit the spectroscopic data for the pion family. We take $m_{u/d}=0.046$ GeV which is the value assumed in hLFQCD together with the Brodsky-de T\'eramond ansatz~\cite{Brodsky:2014yha}. For the transverse confinement scale, we use $\kappa=0.523$ GeV, which is the universal transverse confinement scale across the full hadron spectrum~\cite{Ahmady:2021yzh}. On the other hand, the longitudinal confinement scale $g$ is not universal across the full hadron spectrum~\cite{Ahmady:2021yzh}, and here we take $g=0.109$ GeV. Following \cite{Ahmady:2021lsh,Ahmady:2021yzh}, the meson's parity and charge conjugation quantum numbers are given by:
\begin{equation}
	P=(-1)^{L+1}\; ,\;\; C=(-1)^{L+S+n_\parallel}\; .
	\label{pcrules}
\end{equation} 

\begin{table}
\caption{Quantum numbers and masses of the pion family.}
%\vspace{0.1cm}
\centering
\begin{tabular}{|c c c c c c|}
\hline
$J^{P(C)}$ & Name & $n_\perp$ & $n_\parallel$ & $L$ & $M$ (MeV) \\
\hline
$0^{-}$ & $\pi(140)$ & 0 & 0 & 0 & 134 \\
$0^{-+}$ & $\pi(135)$ & 0 & 0 & 0 & 134 \\
$1^{+-}$ & $b_{1}(1235)$ & 0 & 2 & 1 & 1087 \\
$0^{-+}$ & $\pi(1300)$ & 1 & 2 & 0 &  1087 \\
$2^{-+}$ & $\pi_{2}(1670)$ & 0 & 4 & 2 & 1533 \\
$0^{-+}$ & $\pi(1800)$ & 2 & 4 & 0 & 1533 \\
$2^{-+}$ & $\pi_2(1880)$ & 1 & 6 & 2 & 1875 \\
\hline
\end{tabular}
\label{Spectroscopy}
\end{table}

\begin{figure}[hbt!]
	\begin{center}
		\includegraphics[width=0.9\linewidth]{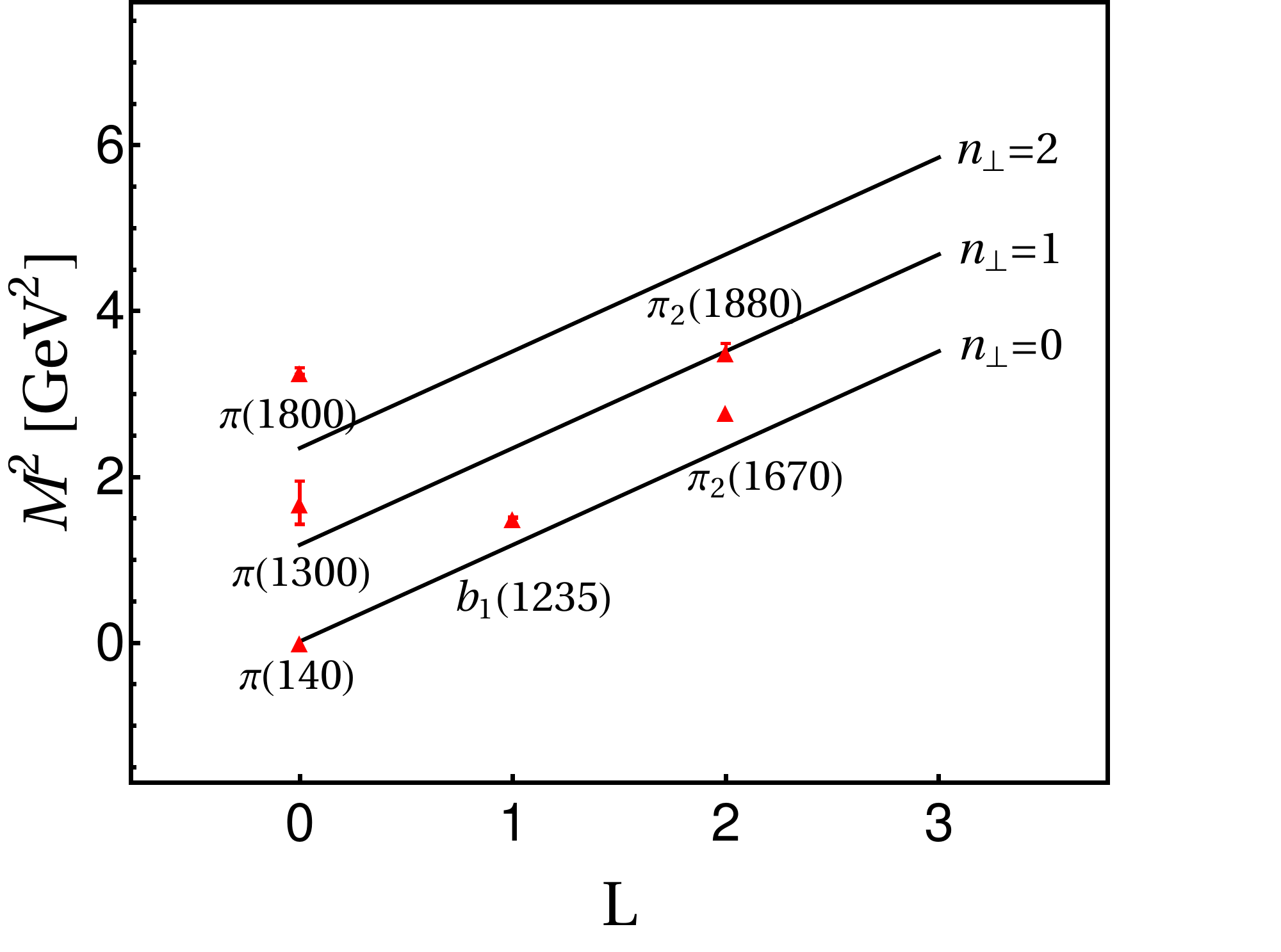}
		\caption{Our Regge trajectories for the pion family using $m_{u/d}=0.046$ GeV,  $g=0.109$ GeV and $\kappa=0.523$ GeV.}
		\label{pionmass}
	\end{center}
\end{figure}

As can be seen in Table~\ref{Spectroscopy}, our computed masses (last column)  are consistent with the measured masses (second column, in parentheses). Notice that the fact that $n_\parallel\ge n_\perp +L$ in Table~\ref{Spectroscopy} is an emerging condition, that is actually observed to remain true across the full hadron spectrum \cite{Ahmady:2021yzh}.  Our resulting Regge trajectories are shown in Fig.~\ref{pionmass}. 

\begin{figure}[hbt!]
	\begin{center}
		\includegraphics[width=0.9\linewidth]{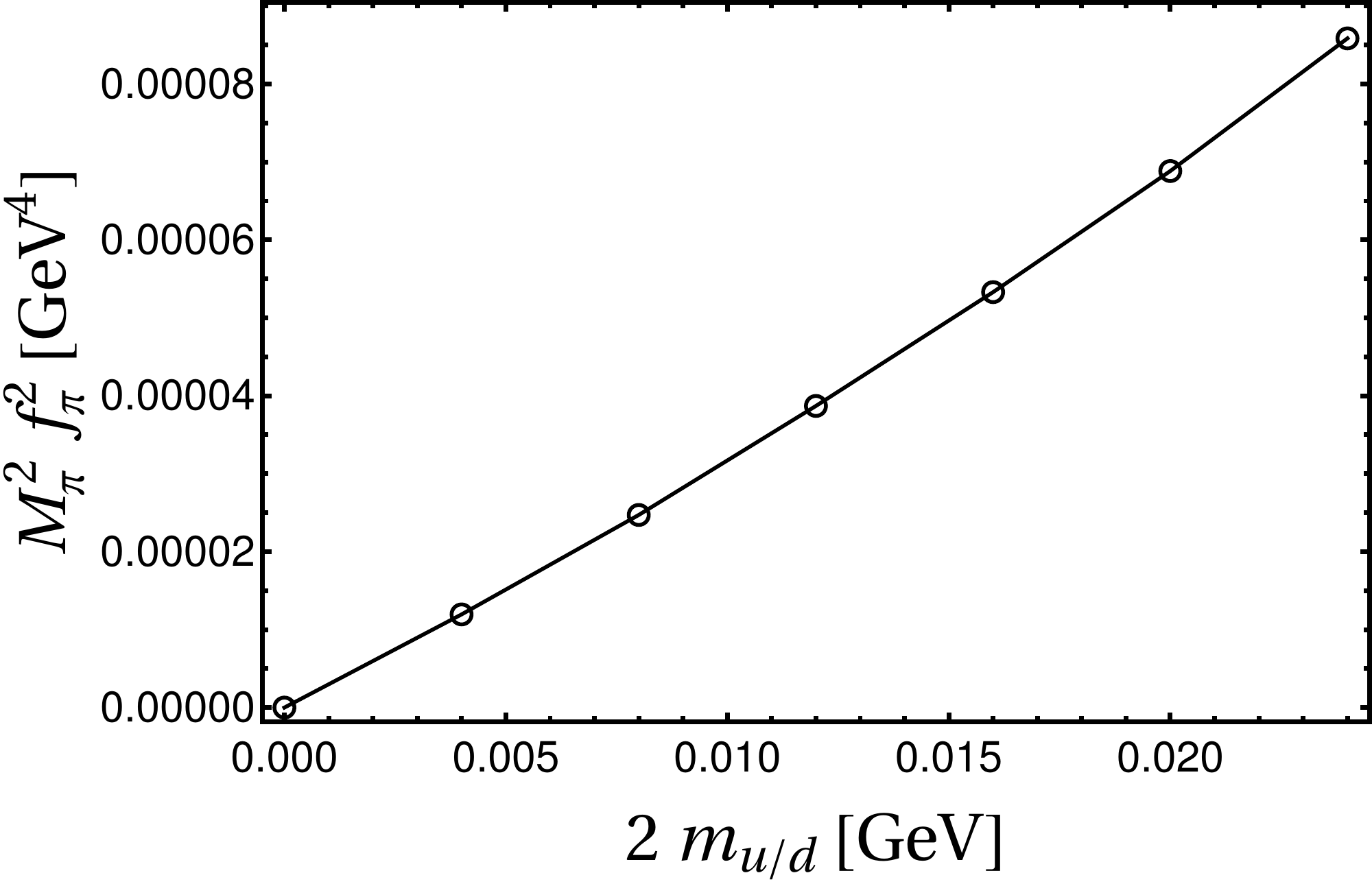}
		\caption{Our numerical result  for $M_\pi^2f_\pi^2$ versus $m_q$ near the chiral limit.  The linear dependence is in accordance with GMOR relation.}
		\label{gmor}
	\end{center}
\end{figure}

It is instructive to check numerically that, as the quark mass goes to zero, our pion mass and decay constant satisfy the GMOR relation given by Eq.~\eqref{GMOR}. This is indeed the case, as shown in Fig.~\ref{gmor}. From the slope of the straight line in Fig.~\ref{gmor}, we can extract a prediction for the in-pion quark condensate: $\langle q\bar{q} \rangle=-0.143~\mathrm{GeV}^{3}$. 
%=========================
%=========================

Having fixed both confinement scales and the light quark mass, we are now in a position to use our pion wavefunction in order to predict various observables. These are parameter-free predictions and provide a stringent test on our proposed pion wavefunction. 
%%%%%%%%%%%%%%%%%%%%%%%%%%%%%%%%%%%%%%%%%
\subsection{Electromagnetic form factor, charge radius, and parton distribution function}
\begin{figure}[hbt!]
	\begin{center}
		\includegraphics[width=0.9\linewidth]{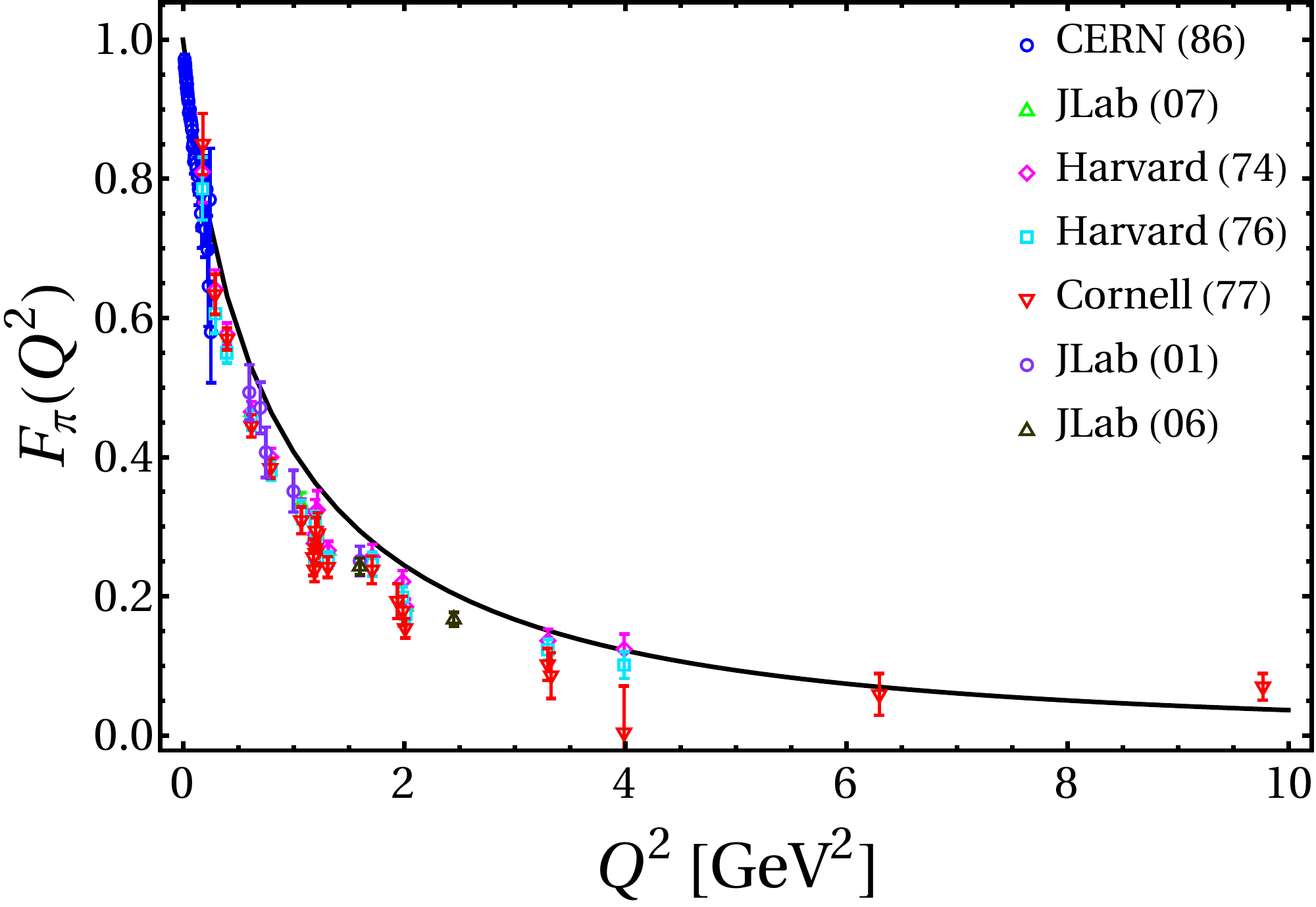}
		\caption{Our prediction (solid-black curve) for pion EM form factor compared with the experimental data~\cite{NA7:1986vav,Bebek:1974iz,Bebek:1977pe,JeffersonLabFpi:2000nlc,CLEO:2005tiu,Seth:2012nn}.}
		\label{emff}
	\end{center}
\end{figure}

%%%%%%%%%%%%%%%%%%%%%%%%%%%%%%%%%%%%%
The pion's electromagnetic (EM) form factor is defined as 
\begin{equation}
	\langle \pi (p^\prime )|J^\mu_{\rm EM} (0)|\pi (p)\rangle =2(p+p^\prime)^\mu F_\pi (Q^2)\; ,
	\label{piffdef}
\end{equation}
where $p^\prime =p+q$ and $Q^2=-q^2$. The quark EM current $J^\mu_{\rm EM}=\sum_f e_f\bar \Psi (z)\gamma^\mu\Psi(z)$ with $f=u,\; d$ and $e_u=+2/3,\; e_d=-1/3$ is the operator responsible for the transition. Using the Drell-Yan-West formula, $F_\pi (Q^2)$ can be written as~\cite{Drell:1969km,West:1970av};
\begin{equation}
	F_\pi (Q^2)=\int \mathrm{d}x{\mathrm{d}}^2\mathbf{b_\perp}J_0[(1-x)b_\perp Q]{|\Psi^\pi (x,\mathbf{b_\perp})|}^2\;.
	\label{ff}
\end{equation} 
Figure~\ref{emff} shows our prediction for $F_\pi (Q^2)$ as compared with the experimental data~\cite{NA7:1986vav,Bebek:1974iz,Bebek:1977pe,JeffersonLabFpi:2000nlc,CLEO:2005tiu,Seth:2012nn}. As can be seen, our prediction is consistent with the data, especially in the low $Q^2$ region, $Q^2 \le 1~\mathrm{GeV}^2$. 

The root-mean-square charge radius of the pion is given by \cite{Brodsky:2007hb}
\begin{align}
	\sqrt{\langle r_\pi^2\rangle}&=\Big[ \frac{3}{2}\int \mathrm{d}x{\mathrm{d}}^2\mathbf{b_\perp}\nonumber\\&\times[b_\perp(1-x)]^2{|\Psi^\pi (x,\mathbf{b_\perp})|}^2\Big]^{1/2}\; .
	\label{radius}
\end{align}
We predict $\sqrt{\langle r_\pi^2\rangle}=0.65$ fm, in excellent agreement with the measured value $0.657 \pm 0.003$ fm~\cite{ParticleDataGroup:2020ssz}.

\begin{figure}[hbt!]
	\begin{center}
		\includegraphics[width=0.9\linewidth]{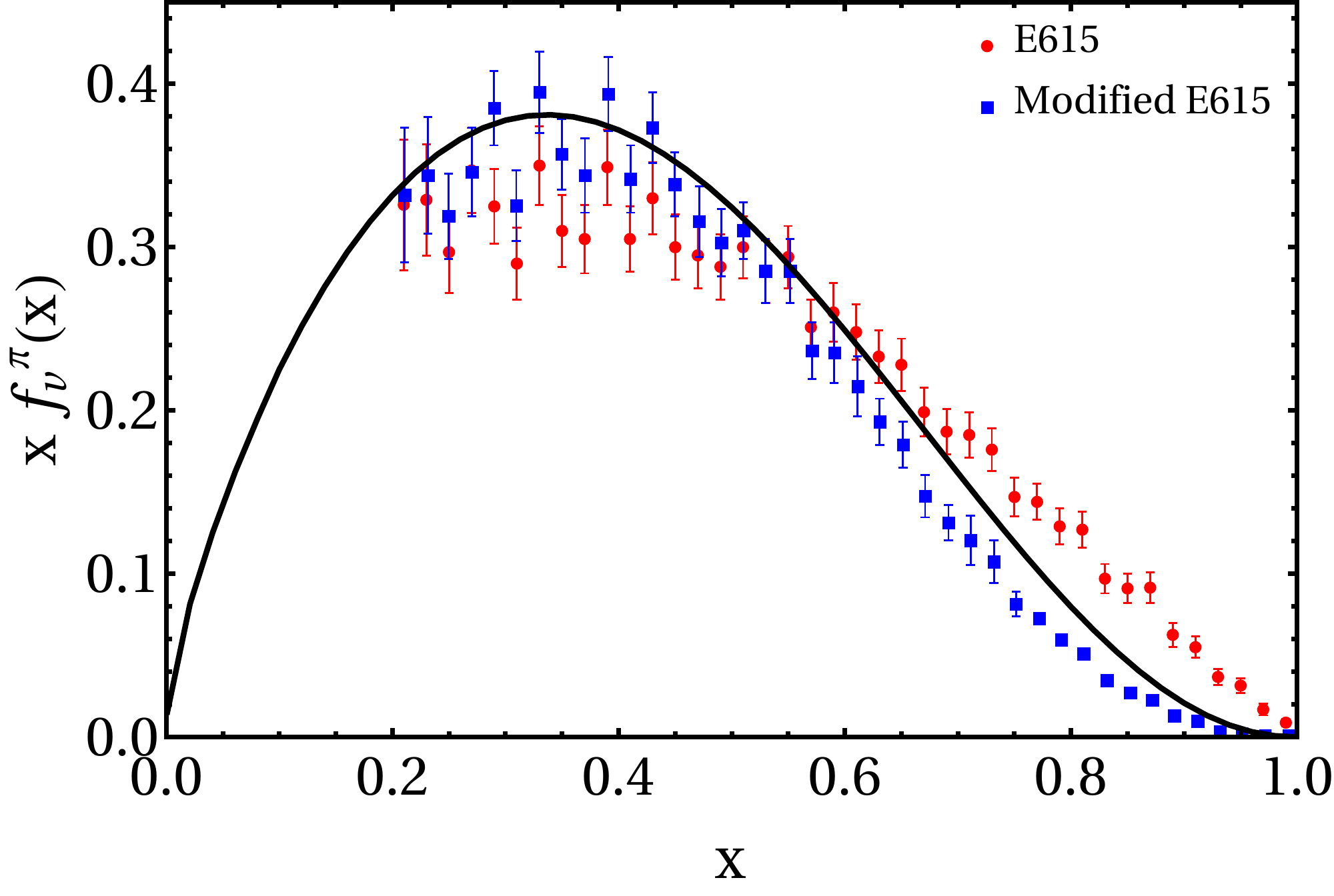}
		\caption{Our prediction (solid-black curve) for pion valence quark PDF compared with the original~\cite{Conway:1989fs} and re-analyzed~\cite{Chen:2016sno} E615 data.}
		\label{fpdf}
	\end{center}
\end{figure}

We also present our prediction for pion PDF, which is defined as:
\begin{equation}
	f_v^\pi (x)=\int {\mathrm{d}}^2\mathbf{b_\perp}{|\Psi^\pi (x,\mathbf{b_\perp})|}^2\;.
	\label{pdf}
\end{equation}
Figure~\ref{fpdf} shows our predicted pion valence quark PDF, which we have evolved from an initial scale $\mu^2_0=0.240~\mathrm{GeV}^2$ to $\mu^2=16~\mathrm{GeV}^2$ relevant to the E615 data using the NNLO DGLAP equations~\cite{Dokshitzer:1977sg,Gribov:1972ri,Altarelli:1977zs} solved numerically using HOPPET~\cite{Salam:2008qg}. Note that we determine the initial scale by matching the first moment of the valence quark PDF to the value extracted, i.e. $2\langle x \rangle_v=0.48 \pm 0.01$, from global fits at $5~\mathrm{GeV}^2$~\cite{Barry:2018ort}. It is interesting to note that our prediction lies in between the original~\cite{Conway:1989fs} and re-analyzed~\cite{Chen:2016sno,Aicher:2010cb} E615 data at large $x$. Specifically, we predict a fall-off like $(1-x)^{2.05} $, consistent with pQCD~\cite{Berger:1979du} and calculations using Dyson-Schwinger Equations~\cite{Hecht:2000xa} as well as the recent analysis by the JLab angular momentum (JAM) collaboration~\cite{Barry:2021osv} and basis light-front quantization (BLFQ) collaboration~\cite{Lan:2021wok}. 

\begin{table}[hbt!]
\caption{Predicted decay constants (DCs) of $\pi$, $\pi^\prime$ and $\pi^{\prime \prime}$ in MeV. The measured value (in MeV) for $f_\pi$ is from Ref.~\cite{ParticleDataGroup:2020ssz}.}
%\vspace{0.1cm}
\centering
\begin{tabular}{|l c c |}
	\hline
	DC & Our results &  Experimental data \\
	\hline
	$f_{\pi}$  & 166.46 & 130.2 $\pm$ 1.7 \\
	$f_{\pi^\prime}$  & 1.44 &  - \\
	$f_{\pi^{\prime\prime}}$ & 0.65 & -  \\
	%\hline
	%$\sqrt{\langle r_\pi^2\rangle}$ [in fm] & 0.66 & $0.657 \pm 0.003$\\
	\hline
\end{tabular}
\label{dcr}
\end{table}

%%%%%%%%%%%%%%%%%%%%%%%%%%%%%%%%%%%
\subsection{Decay constants, distribution amplitude and transition form factor}
%%%%%%%%%%%%%%%%%%%%%%%%%%%%%%%%%%%

The twist-2 pion DA is defined as \cite{PhysRevD.22.2157,Radyushkin:1977gp}
\begin{align}
	&\langle 0 |\bar\Psi_d(z)\gamma^+\gamma_5\Psi_u (0)|\pi^+\rangle\nonumber\\
	& \quad\quad=f_\pi P^+\int \mathrm{d} xe^{ix(P.z)} \phi_\pi (x,\mu )\; ,
	\label{da}
\end{align}
where $z^2=0$ and  $f_\pi$ is pion decay constant given as
\begin{equation}
	f_\pi=\sqrt{\frac{2 N_c}{\pi}} \int \mathrm{d}x  \Psi^\pi(x,{\bf b}_\perp) \big\vert_{{\bf b}_\perp =0}\,.
	\label{dc}
\end{equation}
Our predicted DA is shown in Fig.~\ref{da}. It coincides with the asymptotic form, $6x(1-x)$, and is consistent with the E791 data~\cite{E791:2000xcx}.

%===============================
\begin{figure}[hbt!]
	\begin{center}
	\includegraphics[width=0.9\linewidth]{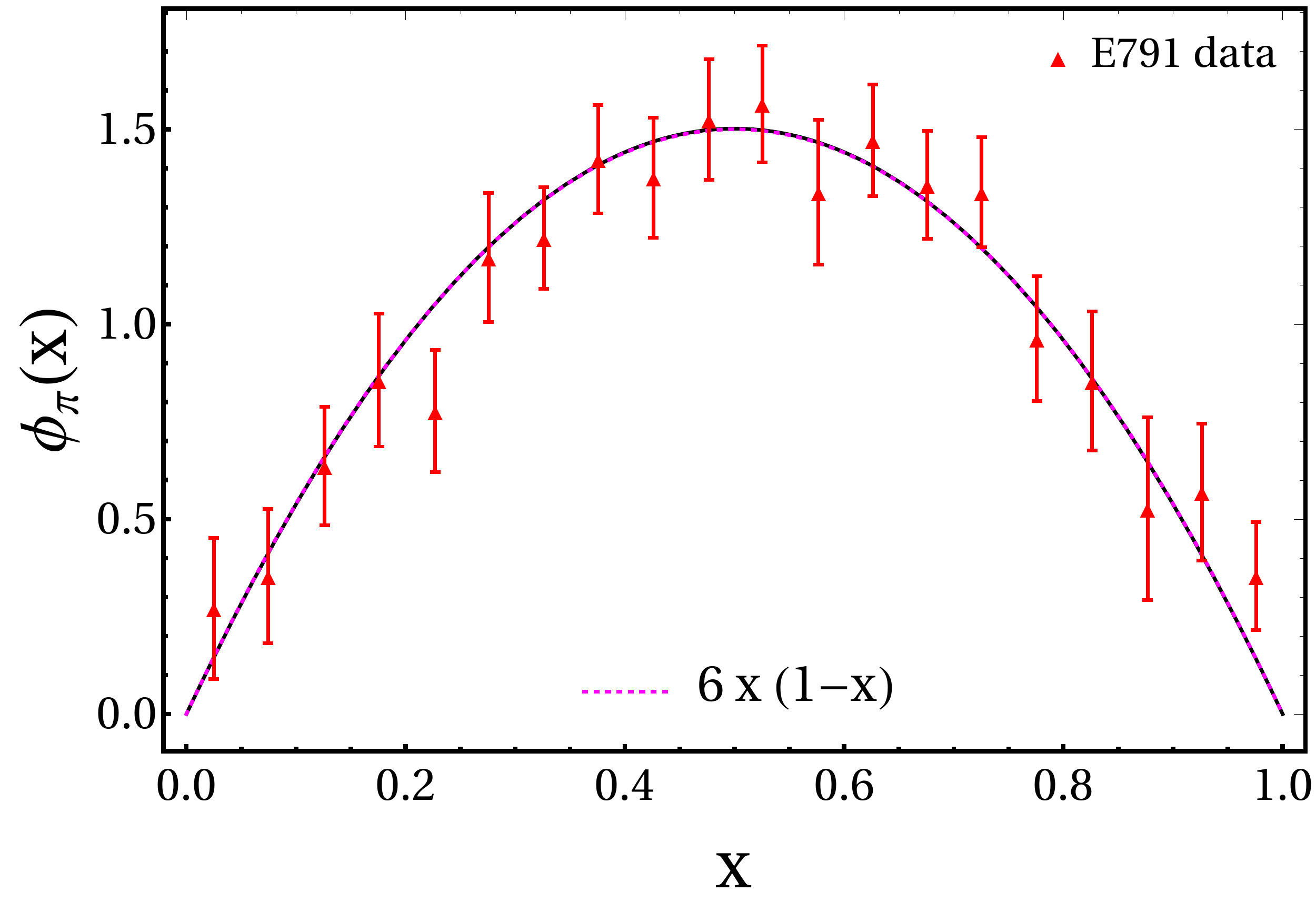}
	\caption{Our prediction (solid-black curve) for pion DA and the asymptotic pion DA (dotted-magenta curve) compared with the E791 data~\cite{E791:2000xcx}.}
	\label{da}
\end{center}
\end{figure}
%================================

As can be seen in Table~\ref{dcr}, our predicted decay constant for pion is in reasonable agreement with the experimental data~\cite{ParticleDataGroup:2020ssz}, while those for the excited states are
suppressed as expected from the GMOR relation. One can associate this suppression to the number of nodes of the longitudinal mode as shown in Fig.~\ref{lm}: the greater the number of nodes, the more is the cancellation between the positive and negative contributions to the right-hand-side of Eq.~\eqref{dc}.

Using the DA given by Eq.~\eqref{da}, we predict the photon-to-pion transition form factor (TFF) given by ~\cite{PhysRevD.22.2157,Mondal:2021czk} 
\begin{align}
Q^2F_{\gamma \pi}(Q^2)=&\frac{\sqrt{2}}{3}f_\pi  \int_0^1 {\rm d}x \,
T_{\mathrm H}(x,Q^2) \nonumber\\
& \times  \phi_\pi (x,(1-x) Q),
\label{eq:TFF_convo}
\end{align} 
where $T_{\mathrm H}(x,Q^2)$ up to the next-to-leading order (NLO) is given by \cite{Braaten:1982yp},
\begin{align}
	&T_{\mathrm H}(x,Q^2)=\frac{1}{1-x} +\frac{\alpha_s(Q^2)}{4 \pi}
C_{\mathrm F} \frac{1}{1-x} \\ & \times \left[-9-\frac{1-x}{x} {\rm ln}(1-x)
+ {\rm ln}^2(1-x)  \right] \nonumber\;.
\label{eq:THNLO}
\end{align}
Here, the color factor is given by $C_{\mathrm F}=\frac{4}{3}$. We use the QCD scale parameter $\Lambda_{\rm QCD}=0.204$ GeV in the strong running coupling $\alpha_s(Q^2)$~\cite{Mondal:2021czk}.

\begin{figure}[hbt!]
	\begin{center}
		\includegraphics[width=0.9\linewidth]{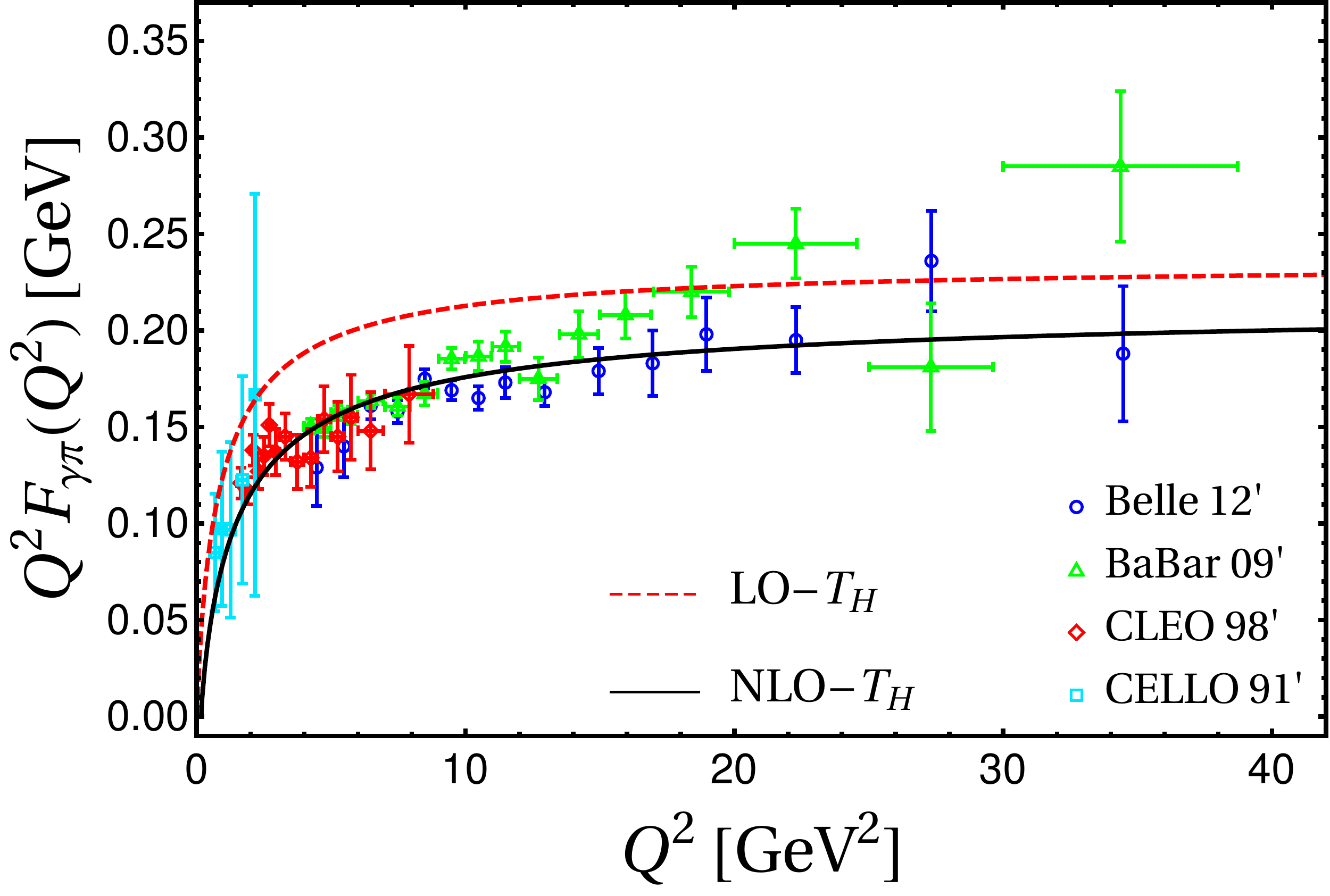}
		\caption{Our prediction (solid-black curve) for photon-to-pion TFF at LO (dashed-red) and NLO (solid-black) compared with the available data \cite{Belle:2012wwz,BaBar:2009rrj,CLEO:1997fho,CELLO:1990klc}.}
		\label{tfffigure}
	\end{center}
\end{figure}

Figure \ref{tfffigure} shows our prediction for $Q^2F_{\gamma \pi}(Q^2)$ to LO and NLO in the hard scattering kernel $T_{\mathrm{H}}$. As we can see, we agree with the CLEO~\cite{CLEO:1997fho}, CELLO~\cite{CELLO:1990klc} and Belle~\cite{Belle:2012wwz} data only in the NLO approximation. Our predictions do not agree well with the rapid rise of the BaBar data~\cite{BaBar:2009rrj}. 

%%%%%%%%%%%%%%%%%%%%%%%%%%%%%%%%%%%%%%%%%%%%%%%%

%%%%%%%%%%%%%%%%%%%%%%%%%%%%%%%%%%%%%%%
\section{Conclusion}
We have shown the holographic Schr\"odinger equation and the 't Hooft equation together capture the three-dimensional internal strong dynamics in the pion and its excited states, leading to a successful simultaneous description of pion spectroscopy and dynamics, while satisfying the chiral-limit constraints implied by the GMOR relation.

\section*{Acknowledgements}
RS and MA are supported by individual Discovery Grants (SAPIN-2020-00051 and SAPIN-2021-00038) from the Natural Sciences and Engineering Research Council of Canada (NSERC). CM thanks the Chinese Academy of Sciences President's International Fellowship Initiative for the support via Grants No. 2021PM0023. CM and SK are supported by new faculty start up funding by the Institute of Modern Physics, Chinese
Academy of Sciences, Grant No. E129952YR0. We thank Sophie Anastasia Tsaltas for cross-checking some of the numerical results in this paper.

\bibliographystyle{elsarticle-num}
\bibliography{ref.bib}
\end{document}